# Terajets produced by 3D dielectric cuboids

V. Pacheco-Peña,[1,a)], M. Beruete,[1,b)] I. V. Minin[2,c)] and O. V. Minin[2,d)]

[1]Antennas Group - TERALAB, Universidad Pública de Navarra, Campus Arrosadía, 31006 Pamplona, Spain

[2]Siberian State Academy of Geodesy, Plahotnogo 10, Novosibirsk, 630108, Russia

The capability of generating *terajets* using 3D dielectric cuboids working at terahertz (THz) frequencies (as analogues of *nanojets* in the infrared band) are introduced and studied numerically. The focusing performance of the terajets are evaluated in terms of the transversal full width at half maximum along *x*- and *y*- directions using different refractive indexes for a 3D dielectric cuboid with a fixed geometry, obtaining a quasi-symmetric terajet with a subwavelength resolution of $\sim 0.46\lambda_0$ when the refractive index is $n = 1.41$. Moreover, the backscattering enhancement produced when metal particles are introduced in the terajet region is demonstrated for a 3D dielectric cuboid and compared with its 2D counterpart. The results of the jet generated for the 3D case are experimentally validated at sub-THz waves, demonstrating the ability to produce terajets using 3D cuboids.

---

[a)] Electronic mail: victor.pacheco@unavarra.es

[b)] Electronic mail: miguel.beruete@unavarra.es (Corresponding author)

[c)] Electronic mail: prof.minin@gmail.com

[d)] Electronic mail: prof.minin@gmail.com

I. **Introduction**

It is known that in order to generate subwavelength resolution in microscopy applications, it is required to deal with the limit imposed by the diffraction of electromagnetic waves.[1] To overcome this problem, different solutions have been reported in the past involving metamaterials[2–4], solid immersion lenses[5], diffractive optics[6,7] and microspherical particles.[8–10]

Several years ago, subwavelength photonic nanojets were introduced and identified using micro scaled cylindrical (2D) and spherical (3D) dielectrics at optical frequencies.[11,12] Photonic nanojets are high intensity, narrow, and non-evanescent beams located at the output surface of a lossless dielectric cylinder/sphere when it is illuminated with a plane wave, demonstrating the ability to obtain a resolution of $\lambda_0/3$ using microspheres, below the diffraction limit.[13,14] Based on this, photonic nanojets have been studied using different properties of the dielectric structure such as non-spherical nanoparticles[15] and also graded index concepts have been applied.[16,17] Besides, experimental demonstration of the backscattering enhancement generated by photonic jets at microwave frequencies has been reported.[18]

In this paper, a new mechanism to produce jets is numerically demonstrated using 3D dielectric cuboids (hexahedrons) working at sub-THZ and THz frequencies. By using a 3D dielectric cuboid with a refractive index $n = 1.41$ immersed in vacuum ($n_0 = 1$), it is shown that "3D terajets" can be obtained just at the surface of the output face of the dielectric structure. Moreover, the performance of the 3D terajets are evaluated when the refractive index of the cuboid is changed from 1.2 up to 2, demonstrating that the position of the focus can be obtained inside or outside the 3D cuboid. Also, the subwavelength focusing properties of 3D dielectric cuboids are numerically studied in terms of the full width at half maximum (FWHM) along both (*x* and *y*) transversal axes, demonstrating a quasi-spherical terajet when $n = 1.41$. Furthermore, the performance of the 3D terajet is evaluated along with its 2D counterpart in terms of the backscattering enhancement perturbation produced when metal microspheres are inserted within the 3D/2D terajets; showing the ability of both structures to enhance the backscattering of metal spheres. Finally, experimental and numerical results of the normalized power distribution along the transversal *x*-axis at $z=0.1\lambda_0$ are performed at sub-THz frequencies using a 3D cuboid of Teflon, demonstrating the ability to produce *"microjets"* using the structure here proposed.

**II.    Terajets performance.**

To begin with, the 3D dielectric cuboid is schematically presented in Fig. 1. The cuboid has lateral dimensions $L = \lambda_0$, along $x$ and $y$ axes while the dimension $H$ is selected to be $1.2\lambda_0$. In order to evaluate the focusing performance of the structure, the transient solver of the commercial software CST Microwave Studio™ is used along with an extra fine hexahedral mesh with a minimum mesh size of $\lambda_0/45$. In the simulation the 3D cuboid is illuminated with a vertically ($E_y$) polarized plane wave (see Fig. 1) at 0.1THz ($\lambda_0 = 3$mm). Moreover, vacuum ($n_0 = 1$) is used as the background medium and expanded open boundary conditions are used in order to insert the 3D cuboid within an infinite medium.

Based on this, the performance of the terajets is evaluated using homogeneous 3D cuboids with different values of refractive index ($n$). Numerical results of the power distribution on the $yz$-plane/$E$-plane (left column) and $xz$-plane/$H$-plane (right column) are shown in Fig. 1 (b-d) for values of $n = 2$, 1.75, 1.41 and 1.2, respectively. Similar to photonic nanojets produced by cylindrical and spherical dielectrics,[11,14] it can be observed that the terajet is located inside the dielectric structure for higher values of refractive index. By decreasing $n$, the terajet is moved away from the surface of the 3D cuboid along the optical axis ($z$) with less power at its focal position. As the refractive index of the cuboid approaches that of the surrounding medium the focusing properties of the 3D cuboid diminish accordingly, as expected.

Moreover, the full width at half maximum (FWHM) along both ($x$ and $y$) transversal directions for positions from 0 up to $1\lambda_0$ along the optical axis ($z$) with a step of $0.2\lambda_0$ are presented in Fig. 2 (a,b) for different values of the 3D cuboid refractive index. Also, the intensity enhancement along the z- axis (calculated as the power distribution with and without the presence of the dielectric cuboids) are shown in Fig. 2(c) for the dielectric cuboids here studied. The best resolution is obtained for $n = 1.75$ with $FWHM_{x,y} < 0.4\lambda_0$ and an intensity enhancement of ~15 times the incident plane wave at the output surface of the 3D cuboid. However, for $n = 1.41$ a quasi-symmetric terajet is obtained with relatively similar values of $FWHM_x$ and $FWHM_y$ and an intensity of ~10 times the power of the incident plane wave. For example, just at the surface of the 3D cuboid, the terajet has a resolution of $FWHM_x = 0.47\lambda_0$ and $FWHM_y = 0.45\lambda_0$. From Fig. 2(c), it can be observed that, even though the intensity enhancement for $n=1.41$ near the cuboid is lower than the value obtained for n=1.75, the intensity enhancement decays smoothly for $n=1.41$, which could be an advantage for applications such as backscattering enhancement[18] in order to detect metal particles away from the surface of the 3D cuboid (see next section). Regarding this last aspect, we can define the parameter *terajet exploration range*, denoted as $\Delta z$, as the distance from the surface at which the intensity enhancement has decayed to half its maximum value (note that we are implicitly assuming here that the maximum appears at the surface, since this is the preferred scenario for microscopy applications). From the figure (dotted lines) it is obvious that the exploration range for $n = 1.41$ is $\Delta z = 0.72\lambda_0$ notably larger than for $n = 1.75$, $\Delta z = 0.16\lambda_0$. Thus, the case $n = 1.41$ is a good tradeoff between intensity enhancement and length of the terajet exploration range. It is important to highlight that for $n = 2$, FWHM values are not represented for all the distances along the optical axis because the terajet is located inside the 3D cuboid and focusing is only observed on the *xz*-plane for distances up to $0.6\lambda_0$. The lower intensity enhancement at the focal position is obtained for the case when $n = 1.2$ (~4) with an exploration range of about $1.96\lambda_0$ because the focus is away from the surface of the dielectric cuboid. Thus, in order to generate jets at the surface of the dielectric cuboid a refractive index contrast relative to the background medium less than 2:1 is required, in agreement with the results of photonic nanojets using circular 2D/3D dielectrics reported recently[11,12]

### III. Backscattering enhancement evaluation.

As it has been described previously, several applications have been reported using photonic nanojets[13]. Similarly to nanojets produced by cylindrical and spherical dielectrics,[11,12,18] here we evaluate the backscattering enhancement when a metal particle is introduced within the terajet with respect to the backscattering without the metal particle.

The evaluation of the backscattering enhancement is numerically studied using the same simulation properties described previously using the 3D cuboid of Fig. 1(d) with a refractive index $n = 1.41$. First a gold sphere ($\sigma_{Au} = 4.561 \cdot 10^7$ S/m) is placed in the vicinity of the output surface of the 3D cuboid and moved along the optical axis (*z*-axis) from $z=0.2\lambda_0$ up to $z=2\lambda_0$ with a step of $0.0833\lambda_0$ [see Fig. 3(a)]. Note that a finite conductivity model is used for the gold sphere, which is a good approximation for frequencies below 10THz. For higher frequencies, a Drude model could fit better the dispersive behavior of metals at infrared frequencies.[19–21] Based on this, the Radar Cross Section (RCS) at the backward direction is obtained with and without the presence of the gold sphere. The backscattering enhancement is calculated as $W_{3D+mp}/W_{3D}$, where $W_{3D+mp}$ is the calculated RCS at the backward direction for the whole system 3D cuboid + metal particle and $W_{3D}$ is the RCS at the backward direction calculated with the 3D cuboid alone.

Simulation results of the backscattering enhancement are presented in Fig. 3(b) when several metal particles with different diameters ($d_1$=0.08$\lambda_0$, $d_2$=0.14$\lambda_0$ and $d_3$=0.2$\lambda_0$) are placed along the optical $z$-axis. It can be observed that the backscattering enhancement changes periodically and the maximum and minimum peaks are located at the same position (0.5$\lambda_0$) along $z$ for the three gold spheres with a periodic oscillation of 0.6$\lambda_0$. Moreover, the maximum values of enhancement are 0.67dB, 2.82dB and 6.44dB for the sphere diameters $d_1$, $d_2$ and $d_3$ respectively. Note that these values are found within the terajet exploration range (which extends up to 0.72$\lambda_0$ along $z$- axis) as expected due to the high power confinement inside this region. For distances far from the terajet, the backscattering is reduced monotonically. Also, in order to evaluate the transversal backscattering enhancement, simulation results are shown in Fig. 3(c) when the gold spheres are placed at $z$=0.5$\lambda_0$ (where the first maximum was obtained) and moved from -0.67$\lambda_0$ to 0.67$\lambda_0$ with a step of 0.16$\lambda_0$ along the $x$-axis. These results demonstrate the ability of the 3D cuboid to produce terajets and enhance the backscattering for different metal spheres along the optical and transversal axes.

For the sake of completeness, the backscattering enhancement of 2D and 3D dielectric cuboids are compared. For the 2D case, the dimension of the cuboid along $y$-axis is changed to 6L (6$\lambda_0$) in order to evaluate its performance as a quasi-infinite structure. Note that by using one dimension larger than the rest of the 2D cuboid, the focusing performance along the $yz$-plane/$E$-plane is changed and a cylindrical Terajet ("*teraknife*") is obtained [see Fig. 3(d-e)]. For both 2D and 3D cases several gold spheres with different diameters ($d_i$) are placed at the distance $z_i = d_i/2$, where i = 1, 2, 3… represents the sphere number, in order to evaluate the backscattering produced by the system when the spheres are touching the output face of the 2D/3D cuboids. Simulation results of the backscattering enhancement are presented in Fig. 3(f) as a function of the diameter of the gold spheres. It is shown that higher values are obtained for the 3D case, in good agreement with previous results[11,12]. This is as expected due to the ability to generate terajets on both $E$- and $H$-planes using 3D cuboids. However, even though lower values of backscattering enhancement are achieved using 2D cuboids, these structures are able to detect particles placed at different positions along the cylindrical *teraknife* by recording its backscattering intensity.

## IV. Experimental results: jet generation

Finally, the existence of terajets is experimentally demonstrated at sub-THz frequencies. As it has been explained previously, the 3D cuboid here proposed can be scaled from microwaves up to optical frequencies. The ability of a 3D cuboid to generate terajets is evaluated in the sub-THZ range at $f_{exp}$=35GHz ($\lambda_{exp}$ = 8.57mm). The dimensions of the 3D cuboid are the same as described in previous sections for the terajet case considering $\lambda_0 = \lambda_{exp}$. The measurements of the power distribution are carried out using the method of movable probe.[7] Experimental and simulation results of the normalized power distribution along the transversal $x$-axis at $z$=0.1$\lambda_0$, i.e., close to the output face of the 3D cuboid of Teflon ($n = 1.46$) are presented in Fig. 4. It is shown that both simulation and experimental results are in good agreement, with a maximum error between them of 7% and a focusing enhancement of ~10 times the incident plane wave. Note that this value is also in good agreement with Fig.2(c) where an enhancement of ~9.5 times the incident plane wave is obtained at the same distance as experiment using a 3D cuboid with $n = 1.41$.

## V. Conclusions

In conclusion, 3D dielectric cuboids working at terahertz frequencies with the ability to generate terajets at its output surface under plane wave illumination have been studied numerically. By changing the refractive index of the 3D cuboids, it has been demonstrated that the focus is moved from inside to outside the structure with a quasi-symmetric terajet and a high intensity of about 10 times the illuminating power when $n$ =1.41. Moreover, it has been demonstrated the ability of such 3D dielectric cuboids to enhance the backscattering perturbation when gold spheres of different sizes are introduced within the terajet region and moved along $z$ and $x$ axis. Finally, the performance of 3D and 2D dielectric cuboids has been studied demonstrating the ability of both structures to enhance the backscattering of gold spheres relative to the backscattering produced by the isolated spheres. The structure here proposed can be used at microwaves and also as a scaled model at optical frequencies, simplifying the fabrication process compared with spherical dielectrics and could find applications in novel microscopy devices.

**ACKNOWLEDGMENTS**

This work was supported in part by the Spanish Government under contract Consolider Engineering Metamaterials CSD2008-00066 and contract TEC2011-28664-C02-01. V.P.-P. is sponsored by Spanish Ministerio de Educación, Cultura y Deporte under grant FPU AP-2012-3796. M.B. is sponsored by the Spanish Government via RYC-2011-08221. In memoriam of our beloved friend Mario Sorolla.

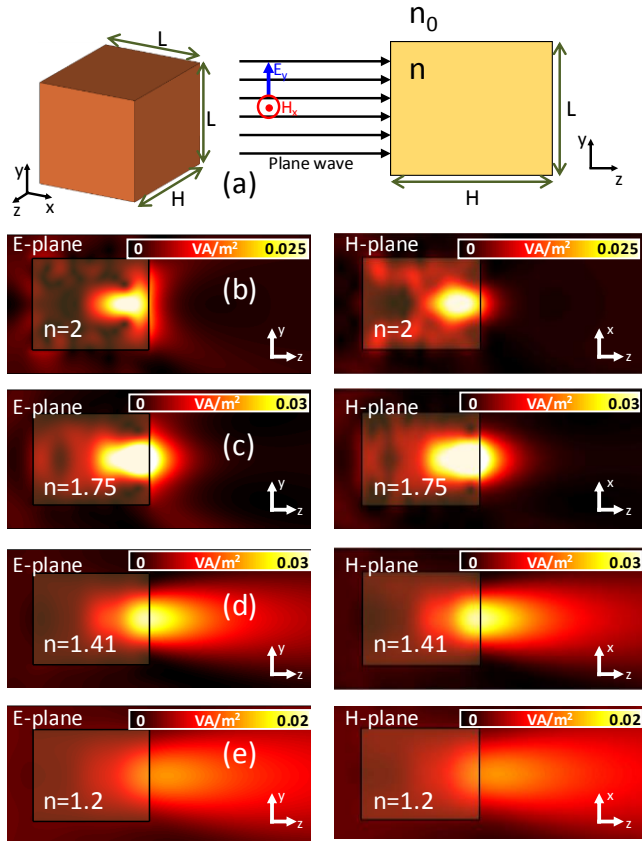

FIG. 1. Schematic representation of the proposed 3D dielectric cuboid with dimensions $L = \lambda_0$ and $H = 1.2\lambda_0$: (a) Perspective and (b) lateral view. Numerical simulations of the terajet performance for different values of the refractive index: (b) $n = 2$, (c) $n = 1.75$, (d) $n = 1.41$ and (e) $n = 1.2$ on the $yz$-plane/$E$-plane (left column) and $xz$-plane/$H$-plane (right column).

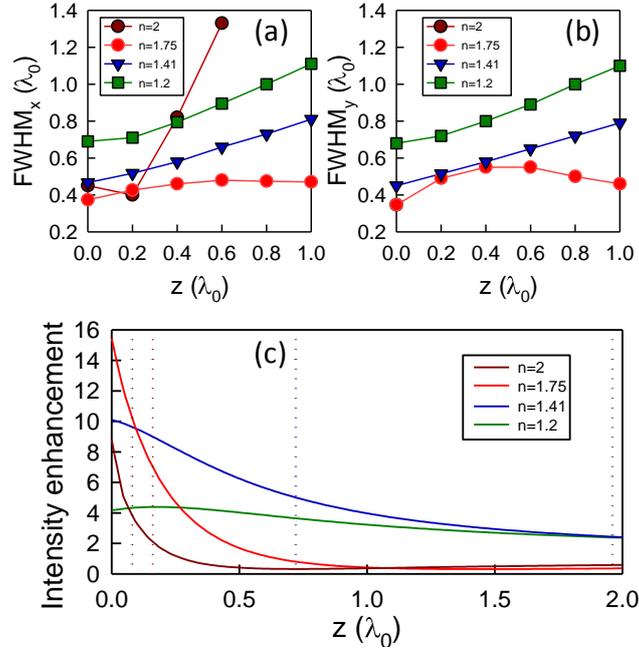

FIG. 2. Full width at half maximum (FWHM) along: (a) *x*-axis and (b) *y*-axis for positions from 0 to $1\lambda_0$ along *z*-axis with a step of $0.2\lambda_0$ for 3D cuboids with refractive index $n = 2$ (brown circles), $n = 1.75$ (red circles), $n = 1.41$ (blue triangles) and $n = 1.2$ (green squares). (c) Intensity enhancement along *z*-axis calculated as the relation between the power distribution with and without the presence of the 3D cuboids of different refractive index. The dotted lines correspond to the exploration range for each jet generated with different refractive indexes.

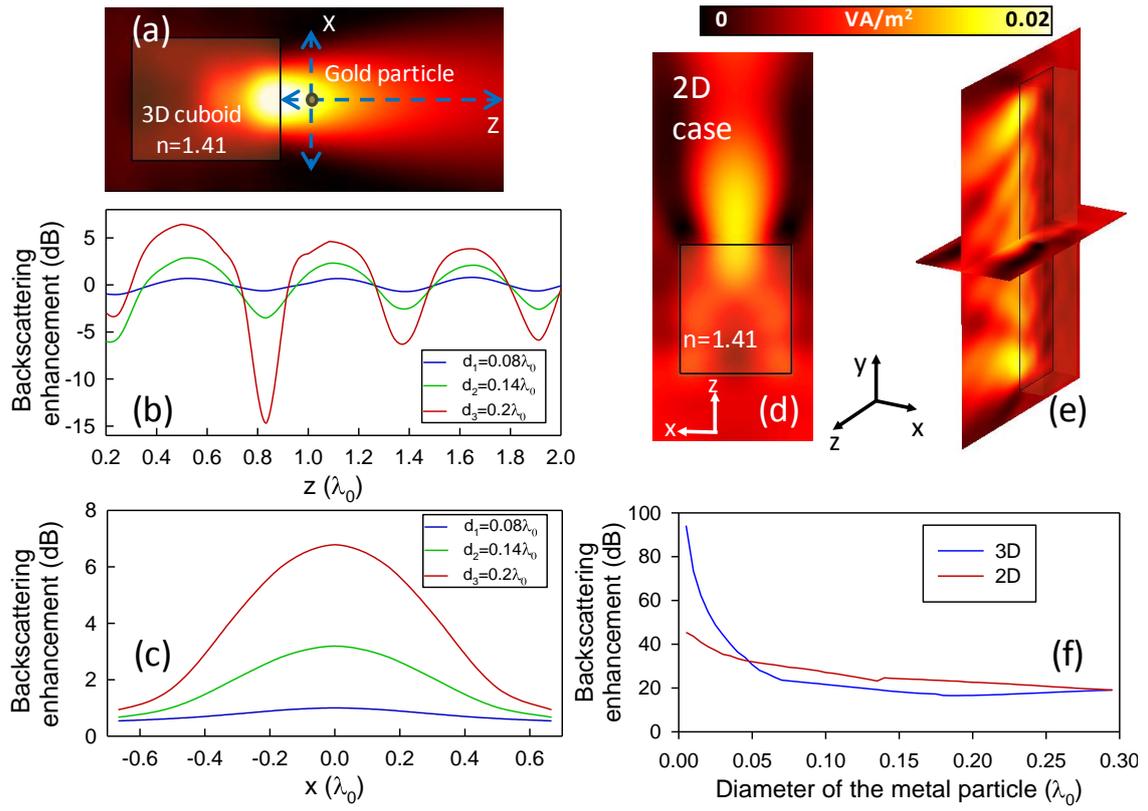

FIG. 3. (a) Schematic representation of the configuration used to evaluate the backscattering enhancement generated by the 3D cuboid. Simulation results of the backscattering enhancement of the system 3Dcuboid+metal particle relative to the 3Dcuboid alone when metal particles with diameters $d_1 = 0.08\lambda_0$ (blue lines), $d_2 = 0.14\lambda_0$ (green lines) and $d_3 = 0.2\lambda_0$ (red lines) are moved along z-axis (b) and along x-axis at $z = 0.5\lambda_0$ (c). Simulation results of the cylindrical Terajet generated for a 2D dielectric cuboid with refractive index $n = 1.41$ with the same dimensions as Figure 1 along x- and z- axes and $6L = 6\lambda_0$ along y-axis: (d) power distribution on the xz-plane/H-plane and (e) on the yz-plane/E-plane. (f) Backscattering enhancement of the system 2D/3D cuboid+metal particle relative to the metal particle alone for different diameters of gold spheres placed at the distance $z_i = d_i/2$ from the output face of the cuboids.

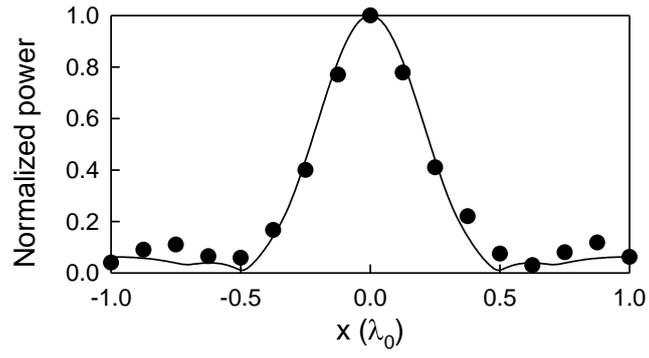

FIG. 4. Experimental (filled circles) and simulation (continuous line) results of the normalized power distribution along the $x$-axis at $z=0.1\lambda_0$.